\documentclass[10pt, twocolumn, pre, aps, superscriptaddress, showpacs]{revtex4-1}
\usepackage{amsmath, graphicx, subfigure, tikz}
\begin{document}

\title{Lower-Critical Dimension of the Random-Field XY Model\\
and the Zero-Temperature Critical Line}

\author{Kutay Ak{\i}n}
    \affiliation{Department of Electrical and Electronics Engineering, Bo\u{g}azi\c{c}i University, Bebek, Istanbul 34342, Turkey}
    \affiliation{Department of Physics, Bo\u{g}azi\c{c}i University, Bebek, Istanbul 34342, Turkey}
\author{A. Nihat Berker}
    \affiliation{Faculty of Engineering and Natural Sciences, Kadir Has University, Cibali, Istanbul 34083, Turkey}
    \affiliation{T\"UBITAK Research Institute for Fundamental Sciences, Gebze, Kocaeli 41470, Turkey}
    \affiliation{Department of Physics, Massachusetts Institute of Technology, Cambridge, Massachusetts 02139, USA}

\begin{abstract}
The random-field XY model is studied in spatial dimensions $d=3$ and 4, and in-between, as the limit $q\rightarrow \infty$ of the $q$-state clock models, by the exact renormalization-group solution of the hierarchical lattice or, equivalently, the Migdal-Kadanoff approximation to the hypercubic lattices.    The lower-critical dimension is determined between $3.81 < d_c <4$.  When the random-field is scaled with $q$, a line segment of zero-temperature criticality is found in $d=3$.  When the random-field is scaled with $q^2$, a universal phase diagram is found at intermediate temperatures in $d=3$.
\end{abstract}
\maketitle
\section{Introduction: Ising and XY Lower-Critical Dimensions}

Quenched randomness strongly affects the occurrence of order at low spatial dimension $d$, reflected as the lower-critical dimension $d_c$ below which no ordering occurs for a given class of systems.  In the random-magnetic-field $n=1$ component spin Ising model, after a strong experimental and theoretical controversy between $d_c=2$ claims \cite{Jaccarino,Wong,BerkerRandH} and $d_c=3$ claims \cite{Birgeneau}, the issue was settled for $d_c=2$.\cite{Machta,Falicov}  The fact that $d_c$ is not 3 fell in contradiction with the prediction of a dimensional shift of 2 due to random fields coming from all-order field-theoretic expansions from $d=6$ down to $d=1$ \cite{Aharony}, which indeed is a considersble distance to expand upon for a small-parameter expansion of $\epsilon=6-d$. In this study, the logically next model, namely the $n=2$ components spin XY model under random fields is examined and surprising results are obtained, this time in near-agreement with the dimensional shift of 2, but also with an interesting zero-temperature critical line segment and a universal scaled finite-temperature phase diagram.

Random-field Ising results supporting $d_c=2$ were obtained \cite{Machta,Falicov} by the Migdal-Kadanoff \cite{Migdal,Kadanoff} renormalization-group calculations in $d=2$ (no random-field order), $d=2.32$ (random-field order), and $d=3$ (more random-field order).  In the same vein, for the random-field XY model, Migdal-Kadanoff renormalization-group calculations are done here in $d=3$ and 4, and in between.  The Migdal-Kadanoff renormalization-group calculation (Fig. 1) is a highly successful, flexible, and therefore most used todate and today, physically motivated approximation for hypercubic lattices and, simultaneously, an exact calculation for $d$-dimensional hierarchical lattices \cite{BerkerOstlund,Kaufman1,Kaufman2}.  The hierarchical lattice connection makes the Migdal-Kadanoff procedure a physically realizable approximation.  For recent work using hierarchical lattices, see Refs. \cite{Derevyagin2,Chio,Teplyaev,Myshlyavtsev,Derevyagin,Shrock,Monthus,Sariyer}. Migdal-Kadanoff-hierarchical-lattices correctly give the lower-critical dimensions of $d_c=1$ of the Ising model \cite{Migdal,Kadanoff}, $d_c=2$ of the XY \cite{Jose,BerkerNelson} and ($n=3$ spin components) Heisenberg \cite{Tunca} models in the absence of quenched randomness.  For the much more complex system with competing quenched-random interactions, Migdal-Kadanoff gives the non-integer $d_c=2.46$ for the Ising spin-glass system.\cite{Parisi,Amoruso,Bouchaud,Boettcher,Demirtas,Parisi2,Atalay}  In addition to giving the lower-critical temperatures, it yields such diverse results as, e.g., the low-temperature algebraic order of the $d=2$ XY model \cite{Jose,BerkerNelson}, the chaotic nature \cite{McKayChaos,McKayChaos2,BerkerMcKay} of the ferromagnetic-antiferromagnetic \cite{Gurleyen} and left-right chiral \cite{Caglar1} Ising spin glasses, and the changeover from second- to first-order phase transitions of $q$-state Potts models in $d=2$ and 3.\cite{Devre}
\begin{figure}[ht!]
\centering
\includegraphics[scale=0.4]{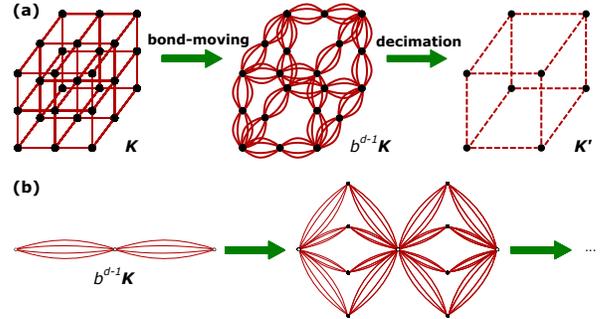}
\caption{From Ref.\cite{Artun}: (a) Migdal-Kadanoff approximate renormalization-group
transformation for the $d=3$ cubic lattice with the length-rescaling
factor of $b=2$. (b) Construction of the $d=3, b=2$ hierarchical
lattice for which the Migdal-Kadanoff recursion relation is exact. For general spatial dimension $d$, the bond-moving is $(b^{d-1})$-fold.  The renormalization-group solution of a hierarchical lattice
proceeds in the opposite direction of its construction.}
\end{figure}

\begin{figure}[ht!]
\centering
\includegraphics[scale=0.2]{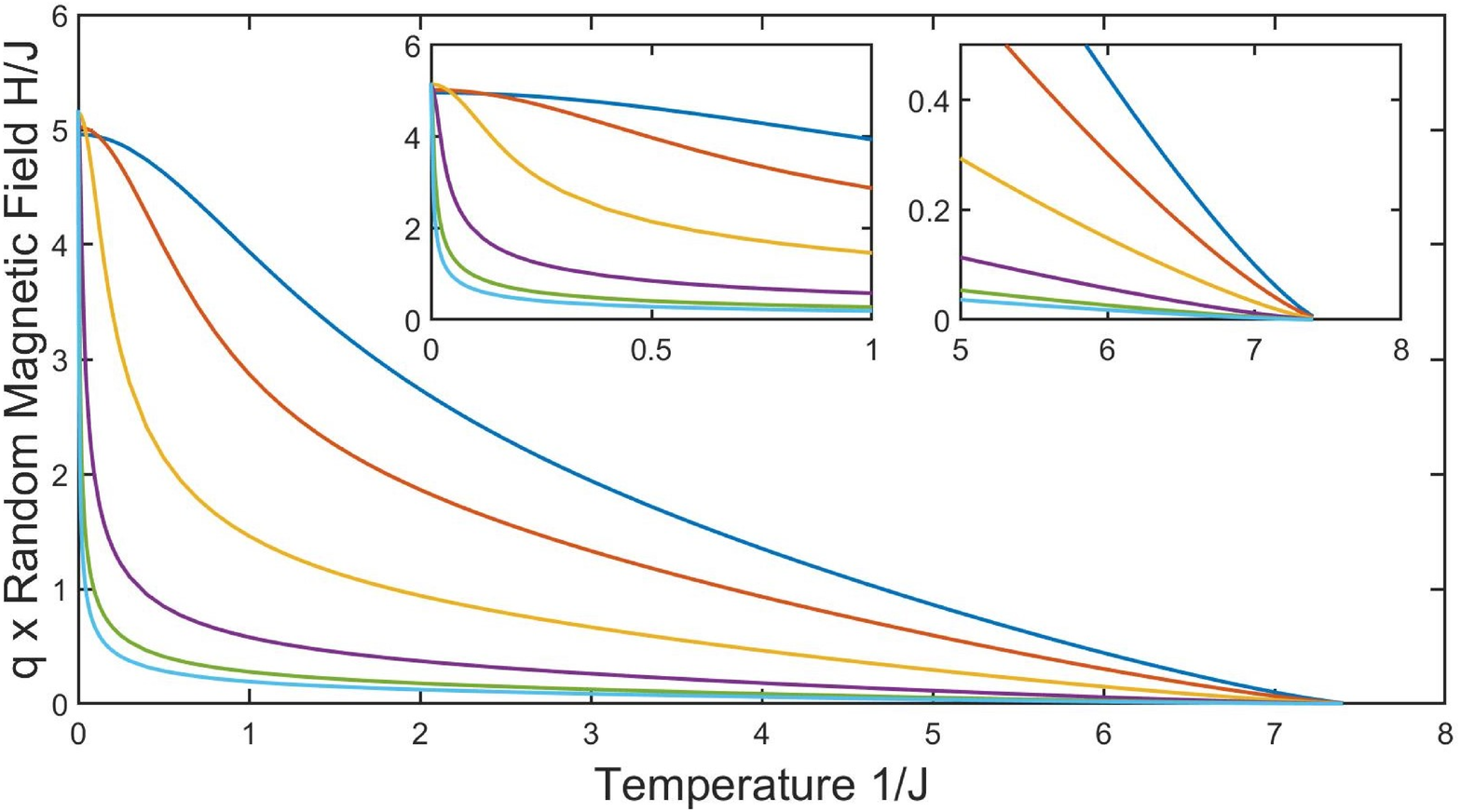}
\caption{Phase diagrams for $(q=7,10,20,50,100,150)$-state random-field clock models in $d=3$, occurring in the figure respectively from high field to low field.  Disordered and ferromagnetic phases occur at high temperature-high field and low temperature-low field, respectively.  It is seen that the ferromagnetic phase, in random field, disappears as $q\rightarrow \infty$, indicating that no ferromagnetic phase occurs in the random-field XY model at non-zero temperature in $d=3$.  However, for high $q$, the ordered phase extends to $qH/J=5.1$ at zero temperature, as also seen in the left box. For high $q$, the zero-field ferromagnetic transition temperature saturates, as also seen in Ref.\cite{Artun} and in the left box in this figure.}
\end{figure}

\begin{figure}[ht!]
\centering
\includegraphics[scale=0.2]{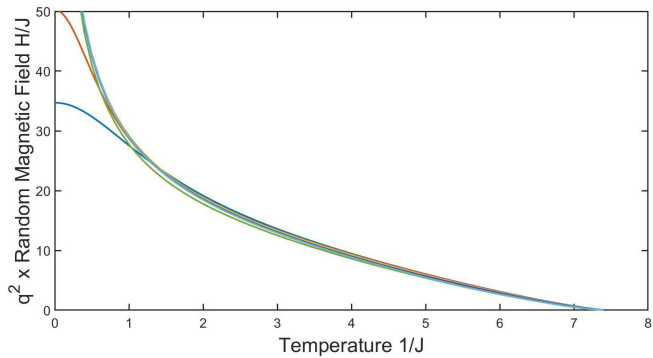}
\caption{Phase diagrams for $(q=7,10,20,50,100,150)$-state random-field clock models in $d=3$. At low temperature, the curves are, from low field to high field, $q=7,10$ and indistinguishably $q=20,50,100,150$.  It is seen that, when the random field is scaled with $q^2$, a universal phase diagram is found above low temperature for high $q$.}
\end{figure}

\section{Model and Method}

The XY model is approached as the $q\rightarrow \infty$ limit of the $q$-state clock models.  In the $q$-state clock models, at each site $i$ of the lattice, a planar unit spin $\vec s_i$ can point in one of $q$ directions in the plane, namely with the angle $\theta_k=k(2\pi/q)$, where $k=0,1,...,q-1$.  A detailed renormalization-group study on the phase transitions and thermodynamics of the $q$-state clock models, without quenched randomness, has been done.\cite{Artun}  The currently studied $q$-state clock model, with quenched random fields, is defined by the Hamiltonian
\begin{equation}
-\beta \mathcal{H}=\sum_{<ij>}{(J\vec s_i\cdot \vec s_j + \vec s_i\cdot \vec H_i + \vec s_j\cdot \vec H_j)},
\end{equation}
where $\beta=1/k_{B}T$ and sum is over all nearest-neighbor pairs of spins.  In each term in the sum, the random-fields $\vec H_i,\vec H_j$ have magnitude $H$ and each randomly points along one of the allowed directions $\theta_k$.

We solve this model using the Migdal-Kadanoff renormalization group.  The local renormalization-group transformation is given in Fig. 1 and is simple to implement in systems without quenched randomness.  With our currently studied quenched random-field model, the renormalization-group evolution of quenched random distributions has to be pursued.  Initially, 5,000 nearest-neighbor Hamiltonians are created, with 10,000 randomly chosen magnetic field directions as described above.  From this distribution, $b^d$ nearest-neighbor Hamiltonians are randomly chosen, to effect the local Migdal-Kadanoff transformation and obtain a renormalized nearest-neighbor Hamiltonian.  This is repeated 5,000 times and the renormalized distribution is obtained.  Each nearest-neighbor Hamiltonian in the distribution is exponentiated and thus kept as a transfer matrix.\cite{Artun,Gurleyen} To conserve, in this distribution, the $(ij) \leftrightarrow (ji)$ and the random-field direction symmetries, each transfer matrix is replicated by its transpose and by the simultaneous cyclic permutations of the rows and columns.  Of the resulting $2q\times 5000$ matrices, 5,000 are randomly chosen. Thus, the distribution continues as 5,000 $q \times q$ matrices.

The flows of the distributions determine the phase diagram:  Renormalization-group trajectories starting in the ferromagnetic phase flow to the strong-coupling sink of $J_{ij}\rightarrow \infty, H_i=0$.  Renormalization-group trajectories starting in the disordered phase flow to the decoupled sink of $J_{ij},H_i=0.$  The boundaries between these flow basins are the phase boundaries.

\section{$d=3$ Dimensions and Zero-Temperature Criticality Segment}

Our calculated phase diagrams for $(q=7,10,20,50,100,150)$-state random-field clock models in $d=3$ are in Fig. 2, occurring in the figure respectively from high field to low field.  Disordered and ferromagnetic phases occur at high temperature-high field and low temperature-low field, respectively.  The $H/J$ values on the vertical axis are multiplied with $q$, originally for better graphical visibility, but eventually leading to a physical result, as seen here.  Firstly, note that the ferromagnetic region under random fields recedes and disappears as $q$ is increased.  This result is even more evident, when we recall that the vertical axis values are amplified by a factor of $q$ for better pictorial visibility.  The ferromagnetic phase, in random field, disappearing as $q\rightarrow \infty$ indicates that no ferromagnetic phase occurs in the random-field XY model at non-zero temperature in $d=3$.

Secondly and quite interestingly, given our choice of vertical axis values, it revealed that the ordered phase extends at very low temperatures, for the high $q$ to the universal value of $qH/J=5.1$.  This is more visible in the left inset box of Fig. 2.  Thus, at $q\rightarrow \infty$, a line segment of zero-temperature critical points occurs between $qH/J=0$ and $qH/J=5.1$.  Zero-temperature critical segments and multicritical points have been found before, under exact renormalization-group treatment, in the $d=1$ Blume-Emery-Griffiths model \cite{Furman}.

Thirdly, for high $q$, the zero-field ferromagnetic transition temperature saturates, as also seen in Ref.\cite{Artun} and in detail in the right inset box in Fig. 2.  Furthermore, when the vertical axis is scaled, not by $q$, but by $q^2$, a universal phase diagram emerges above low temperature for high $q$, as seen in Fig. 3.

\begin{figure}[ht!]
\centering
\includegraphics[scale=0.2]{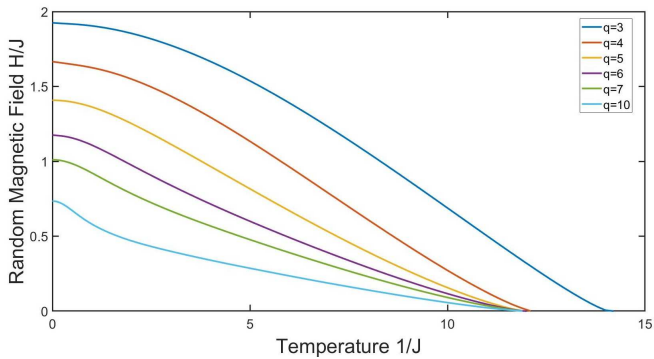}
\caption{Phase diagrams for $(q=3,4,5,6,7,10)$-state random-field clock models in $d=3.32$, occurring in the figure respectively from high field to low field.}
\end{figure}

\begin{figure}[ht!]
\centering
\includegraphics[scale=0.2]{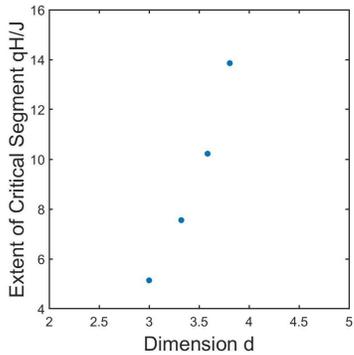}
\caption{The critical line segment, at zero temperature, is between $qH/J=0$ and the $qH/J$ values shown in this figure for each dimension $d$.  The values are consistent with a divergence as $d=4$ is approached.}
\end{figure}

\begin{figure}[ht!]
\centering
\includegraphics[scale=0.2]{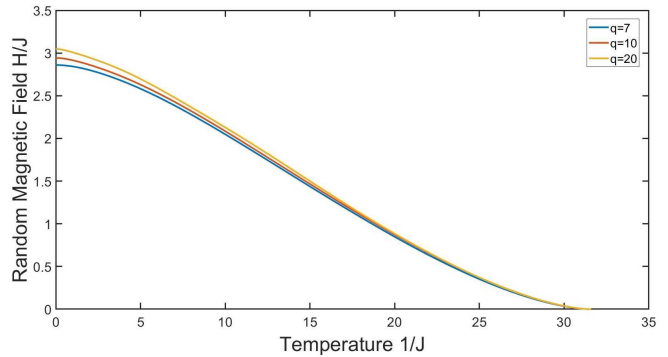}
\caption{Phase diagrams for $(q=7,10,20)$-state random-field clock models in $d=4$, occurring in the figure respectively from low field to high field.}
\end{figure}

\section{$d=4$ Dimensions and the Lower-Critical Dimension}

The phase diagrams for $(q=3,4,5,6,7,10)$-state random-field clock models in $d=3.32$ are shown in Fig. 4.  It is again seen that the ferrromagnetic phase, under random fields, recedes and disappears as $q\rightarrow \infty$.  Thus, no ferromagnetic phase occurs under random fields in the XY model in $d=3.32$.  However, our calculation again gives the zero-temperature critical segment, between $qH/J=0$ and $qH/J=7.6$ universally for all $q$ in $d=3.32$.

The same results are obtained for $d=3.58$ and $3.81$, with the zero-temperature critical segment expanding, reaching $qH/J=10.2$ and 13.9, respectively.

A qualitatively different picture occurs in the phase diagrams for $d=4$, seen in Fig. 5.  Going from $q=7$ to $q=10$, the ferromagnetic phase slightly expands in the random field, as opposed to drastically receding as in the lower dimensions. Going from $q=10$ to $q=20$, a much larger $q$ interval, the ferromagnetic phase even more slightly expands in the random field.  Thus, the ferromagnetic phase occurs, under random fields, for $q\rightarrow \infty$ and for the XY model in $d=4$.

We thus see that the lower-critical dimension for the random-field XY model is between $d=3.81$ and $d=4$, namely $3.81 < d_c <4$.

\section{Conclusion}

In order to investigate the random-field XY model, we have studied the random-field $q$-state clock models for increasing $q$, for dimensions $d=3,3.32,3.58,3.81,4$.  We find that for the random-field XY model, the lower-critical dimension is between $d=3.81$ and $d=4$, namely $3.81 < d_c <4$.  At $d<d_c$, we find a zero-temperature segment of criticality, stretching from zero to a value of $qH/J$ that is $q$-independent for large $q$ and that increases as $d_c$ is approached.

\begin{acknowledgments}
Support by the Academy of Sciences of Turkey (T\"UBA) is gratefully acknowledged.
\end{acknowledgments}


\begin{references}

\bibitem{Jaccarino} D. P. Belanger, A. R. King, and V. Jaccarino, Random-field effects on critical behavior of diluted Ising antiferromagnets, Phys. Rev. Lett. {\bf 48}, 1050 (1982).
\bibitem{Wong} P.-Z. Wong and J. W. Cable, Hysteretic behavior of the diluted random-field Ising system $Fe_{0.70}Mg_{0.30}Cl_2$, Phys. Rev. B {\bf 28}, 5361 (1983).
\bibitem{BerkerRandH} A. N. Berker, Ordering under random fields: Renormalization-group arguments, Phys. Rev. B {\bf 29}, 5243 (1984).
\bibitem{Birgeneau} H. Yoshizawa, R. A. Cowley, G. Shirane, R. J. Birgeneau, H. J. Guggenheim, and H. Ikeda, Random-field effects in two- and three-dimensional Ising antiferromagnets, Phys. Rev. Lett. {\bf 48}, 438 (1982).
\bibitem{Machta} M. S. Cao and J. Machta, Migdal-Kadanoff study of the random-field Ising model, Phys. Rev. B {\bf 48}, 3177 (1993).
\bibitem{Falicov} A. Falicov, A. N. Berker, and S. R. McKay, Renormalization-group theory of the random-field Ising model in 3 dimensions, Phys. Rev. B {\bf 51}, 8266 (1995).
\bibitem{Aharony} A. Aharony, Y. Imry, and S.-k. Ma, Lowering of dimensionality in phase transitions with random fields, Phys. Rev. Lett. {\bf
    37}, 1364 (1976).

\bibitem{Migdal} A. A. Migdal, Phase transitions in gauge and spin lattice systems, Zh. Eksp. Teor. Fiz. {\bf69}, 1457 (1975) [Sov. Phys. JETP {\bf42}, 743 (1976)].
\bibitem{Kadanoff} L. P. Kadanoff, Notes on Migdal's recursion formulas, Ann. Phys. (N.Y.) {\bf100}, 359 (1976).
\bibitem{BerkerOstlund} A. N. Berker and S. Ostlund, Renormalisation-group calculations of finite systems: Order parameter and specific heat for epitaxial ordering, J. Phys. C {\bf 12}, 4961 (1979).
\bibitem{Kaufman1} R. B. Griffiths and M. Kaufman, Spin systems on hierarchical lattices: Introduction and thermodynamic Limit, Phys. Rev. B {\bf 26}, 5022R (1982).
\bibitem{Kaufman2} M. Kaufman and R. B. Griffiths, Spin systems on hierarchical lattices: 2. Some examples of soluble models, Phys. Rev. B {\bf 30}, 244 (1984).

\bibitem{Jiang} K. Jiang, J. Qiao, and Y. Lan, Chaotic renormalization flow in the Potts model induced by long-range competition, Phys. Rev. E {\bf 103}, 062117 (2021).
\bibitem{Derevyagin2} G. Mograby, M. Derevyagin, G. V. Dunne,  and A. Teplyaev, Spectra of perfect state transfer Hamiltonians on fractal-like graphs, J. Phys. A {\bf 54}, 125301 (2021).
\bibitem{Chio} I. Chio, R. K. W. Roeder, Chromatic zeros on hierarchical lattices and equidistribution on parameter space, Annales de l'Institut Henri Poincar\'{e} D, {\bf 8}, 491 (2021).
\bibitem{Teplyaev} B. Steinhurst and A. Teplyaev, Spectral analysis on Barlow and Evans’ projective limit fractals, J. Spectr. Theory {\bf 11}, 91 (2021).
\bibitem{Myshlyavtsev} A. V. Myshlyavtsev, M. D. Myshlyavtseva, and S. S. Akimenko, Classical lattice models with single-node interactions on hierarchical lattices: The two-layer Ising model, Physica A {\bf 558}, 124919 (2020).
\bibitem{Derevyagin} M. Derevyagin, G. V. Dunne, G. Mograby, and A. Teplyaev, Perfect quantum state transfer on diamond fractal graphs, quantum information processing, {\bf19}, 328 (2020).
\bibitem{Shrock} S.-C. Chang, R. K. W. Roeder, and R. Shrock, q-Plane zeros of the Potts partition function on diamond hierarchical graphs, J. Math. Phys. {\bf61}, 073301 (2020).
\bibitem{Monthus} C. Monthus, Real-space renormalization for disordered systems at the level of large deviations, J. Stat. Mech. - Theory and Experiment, 013301 (2020).
\bibitem{Sariyer} O. S. Sar{\i}yer, Two-dimensional quantum-spin-1/2 XXZ magnet in zero magnetic field: Global thermodynamics from renormalisation group theory, Philos. Mag. {\bf 99}, 1787 (2019).

\bibitem{Jose} J. V. Jos\'{e}, L. P. Kadanoff, S. Kirkpatrick, and D. R. Nelson, Renormalization, vortices, and symmetry-breaking perturbations in 2-dimensional planar Model, Phys. Rev. B {\bf16}, 1217 (1977).
\bibitem{BerkerNelson} A. N. Berker and D. R. Nelson, Superfluidity and phase separation in Helium films, Phys. Rev. B {\bf 19}, 2488 (1979).
\bibitem{Tunca} E. Tunca and A. N. Berker, Renormalization-group theory of the Heisenberg model in d dimensions, arXiv:2202.06049 [cond-mat.stat-mech] (2022).

\bibitem{Parisi} S. Franz, G. Parisi, and M.A. Virasoro, Interfaces and lower critical dimension in a spin-glass model, J. Physique I {\bf 4}, 1657 (1994).
\bibitem{Amoruso} C. Amoruso, E. Marinari, O. C. Martin, and A. Pagnani, Scalings of domain wall energies in two dimensional Ising spin glasses, Phys. Rev. Lett. {\bf 91}, 087201 (2003).
\bibitem{Bouchaud} J.-P. Bouchaud, F. Krzakala, and O. C. Martin, Energy exponents and corrections to scaling in Ising spin glasses, Phys. Rev. B {\bf 68}, 224404 (2003).
\bibitem{Boettcher} S. Boettcher, Stiffness of the Edwards-Anderson model in all dimensions, Phys. Rev. Lett. {\bf95}, 197205 (2005).
\bibitem{Demirtas} M. Demirtaş, A. Tuncer, and A. N. Berker, Lower-critical spin-glass dimension from 23 sequenced hierarchical models, Phys. Rev. E 92, 022136 (2015).
\bibitem{Parisi2} A. Maiorano and G. Parisi, Support for the value 5/2 for the spin glass lower critical dimension at zero magnetic field, Proc. Natl. Acad. Sci. USA {\bf 115}, 5129 (2018).
\bibitem{Atalay} B. Atalay and A. N. Berker, A lower lower-critical spin-glass dimension from quenched mixed-spatial-dimensional spin glasses, Phys. Rev. E {\bf 98}, 042125 (2018).

\bibitem{McKayChaos} S. R. McKay, A. N. Berker, and S. Kirkpatrick, Spin-glass behavior in frustrated Ising models with chaotic renormalization-group trajectories, Phys. Rev. Lett. {\bf 48}, 767 (1982).
\bibitem{McKayChaos2} S. R. McKay, A. N. Berker, and S. Kirkpatrick, Amorphously packed, frustrated hierarchical models: Chaotic rescaling and spin-glass behavior, J. Appl. Phys. {\bf 53}, 7974 (1982).
\bibitem{BerkerMcKay} A. N. Berker and S. R. McKay, Hierarchical models and chaotic spin glasses, J. Stat. Phys. {\bf 36}, 787 (1984).
\bibitem{Gurleyen} S.E. G\"urleyen and A.N. Berker, Asymmetric phase diagrams, algebraically ordered BKT phase, and peninsular Potts flow structure in long-range spin glasses, S.E. G\"urleyen and A.N. Berker, Phys. Rev. E {\bf 105}, 024122 (2022).
\bibitem{Caglar1} T. \c{C}a\u{g}lar and A. N. Berker, Chiral Potts spin glass in d = 2 and 3 dimensions, Phys. Rev. E {\bf 94}, 032121 (2016).
\bibitem{Devre} H. Y. Devre and A. N. Berker, First-order to second-order phase transition changeover and latent heats of q-state Potts models in d=2,3 from a simple Migdal-Kadanoff adaptation, arXiv:2202.01528 [cond-mat.stat-mech] (2022).

\bibitem{Artun} E. C. Artun and A. N. Berker, Complete density calculations of q-state Potts and clock models: Reentrance of interface densities under symmetry breaking, Phys. Rev. E {\bf 102}, 062135 (2020).

\bibitem{Furman} S. Krinsky and D. Furman, Exact renormalization group exhibiting tricritical fixed point for a spin-one Ising model in one dimension, Phys. Rev. B {\bf 11}, 2602 (1975).

\end{references}
\end{document}